\documentclass[aps,prd,twocolumn,nofootinbib,superscriptaddress,floatfix]{revtex4-2}
\pdfoutput=1
\usepackage{amsmath,amssymb,bm,graphicx,booktabs,xcolor,hyperref}
\usepackage[T1]{fontenc}
\hypersetup{colorlinks=true,citecolor=blue,linkcolor=blue,urlcolor=blue}
\newcommand{\TeV}{\,\mathrm{TeV}}
\newcommand{\GeV}{\,\mathrm{GeV}}
\newcommand{\keV}{\,\mathrm{keV}}
\newcommand{\MeV}{\,\mathrm{MeV}}
\newcommand{\vthree}{v_{331}}

\newcommand{\RXL}{R_{XL}}
\newcommand{\dmzero}{\delta m_0}
\newcommand{\whI}{\widehat I_1}
\begin{document}

\title{LZ-Motivated Pseudo-Dirac Higgsinos in the Supersymmetric 331 Model from the Supersymmetric SU(6) GUT Model}

\author{Imtiaz Khan}
\email{ikhanphys1993@gmail.com}
\affiliation{Department of Physics, Zhejiang Normal University, Jinhua, Zhejiang 321004, China}
\affiliation{Research Center of Astrophysics and Cosmology, Khazar University, Baku, AZ1096, 41 Mehseti Street, Azerbaijan}

\author{Ali Muhammad}
\email{alimuhammad@phys.qau.edu.pk}
\affiliation{CAS Key Laboratory of Theoretical Physics, Institute of Theoretical Physics, Chinese Academy of Sciences, Beijing 100190, China}
\affiliation{School of Physical Sciences, University of Chinese Academy of Sciences, No. 19A Yuquan Road, Beijing 100049, China}

\author{G. Mustafa}
\email{gmustafa3828@gmail.com}
\affiliation{Department of Physics, Zhejiang Normal University, Jinhua, Zhejiang 321004, China}

\author{Farruh~Atamurotov}
\email{atamurotov@yahoo.com}
\affiliation{Kimyo International University in Tashkent, Shota Rustaveli str. 156, Tashkent 100121, Uzbekistan}

\author{Ahmadjon~Abdujabbarov}
\email{ahmadjonab@gmail.com}
\affiliation{School of Physics, Harbin Institute of Technology, Harbin 150001, People's Republic of China}
\author{Mussawir Khan} 
\email{mussawirkhan@ihep.ac.cn}
\affiliation{State Key Laboratory of Particle Astrophysics, Institute of High Energy Physics, Chinese Academy of Sciences, Beijing 100049, China}
\affiliation{University of Chinese Academy of Sciences, Beijing 100049, China}

\begin{abstract}
We study the realization of a sub-MeV pseudo-Dirac Higgsino mass splitting, motivated by the high-recoil candidate reported by the LUX--ZEPLIN (LZ) experiment, within a supersymmetric $SU(6)$ grand unified framework with an intermediate $SU(3)_C\times SU(3)_L\times U(1)_X$ gauge symmetry. Unlike conventional MSSM constructions, hypercharge emerges only after the $331$ symmetry breaking, and the cancellation
responsible for a small Higgsino splitting must therefore be formulated in terms of the underlying $U(1)_X$ and $SU(3)_L$ gaugino invariants. We derive the corresponding
leading-order relation, $I_X=-I_L$, and identify its correlation with the colored gaugino invariant for a singlet-adjoint contribution to the gauge kinetic function. To investigate its realization, we diagonalize the complete
tree-level neutral gauge-Higgs fermion sector, comprising the hypercharge-connected ten-state block and a complementary six-state block. The resulting spectrum exhibits continuous
sub-MeV matching solutions, including a $350\,\mathrm{keV}$ reference splitting for a predominantly Higgsino state with
$|\mu|=1.091\,\mathrm{TeV}$. We further construct a symmetry-compatible higher-dimensional messenger-splitting
operator involving an additional singlet spurion. This realization correlates the colored and electroweak gaugino masses while requiring an adjusted messenger mass ratio. We assess its implications for gauge-mediated scalar masses,
the stability of the $331$ vacuum, gravitino mass ordering, and conditional two-loop gauge coupling unification with split messenger thresholds. Finally, the leading-order
Higgsino coannihilation calculation yields
$\Omega_\chi h^2\simeq 0.119$ at the reference mass. However, the nearly unsuppressed off-diagonal $Z$ coupling of the light neutral states makes the $350\,\mathrm{keV}$
solution incompatible with standard full-density solar-capture constraints. Our results provide a conditional ultraviolet realization of sub-MeV Higgsino splitting while identifying
the messenger-sector assumptions, threshold sensitivities, and phenomenological limitations of this scenario. The splitting remains a tree-level matching result; a renormalized pole-mass prediction requires a dedicated sequential threshold and self-energy calculation.
\end{abstract}
\maketitle

\section{Introduction}\label{sec:intro}
The LUX--ZEPLIN (LZ) collaboration has extended its nuclear-recoil analysis to approximately $270\keV$ and reported one event at $248\pm23_{\rm stat}\pm23_{\rm sys}\keV$ in a region with a low expected background. The maximum local significance is $3.4\sigma$ and the global significance is $2.6\sigma$ after the look-elsewhere effect~\cite{LZ2026}. Endothermic scattering gives a direct reason for considering this recoil range: the excitation energy suppresses low-energy recoils and selects the high-velocity tail of the incident dark-matter distribution~\cite{TuckerSmithWeiner2001,Bramante2016}.

A nearly pure Higgsino provides a simple electroweak realization. Its two neutral Weyl fields form a pseudo-Dirac pair when the electroweak gauginos decouple, while Majorana gaugino masses generate the neutral splitting. The LZ-motivated analyses place the relevant splitting in the few-hundred-keV range for a Higgsino near the thermal mass~\cite{FanReece2026,FreeseTheodosopoulos2026,WuZhangZhu2026,DiMauro2026}. This small gap is not generic in the MSSM. Away from cancellations it points to a very high electroweak-gaugino scale, whereas an opposite-sign bino--wino contribution can reduce the gap at much lower masses~\cite{NagataShirai2015,ChunJungPark2017,DuWang2026}.

Several recent studies have examined different aspects of the same event. The $SU(5)$ construction of Ref.~\cite{DuWang2026} realizes the bino--wino cancellation through a nonuniversal GUT boundary. The GNMSSM instead uses Higgsino--singlino mixing~\cite{BisalCaoLi2026}, and a separate $SU(6)$-based GUT construction correlates a pseudo-Dirac splitting with a heavy $Z'$~\cite{KotlarskiKowalskaSessolo2026}. The high-energy sideband and annual modulation provide additional tests of endothermic interpretations~\cite{RoddSafdiSlatyerXu2026,McCabe2026}. Solar capture is restrictive for a Higgsino because the off-diagonal $Z$ current is nearly unsuppressed; current analyses place the relevant lower splitting near the $0.5$--$0.6\MeV$ range under their stated assumptions~\cite{PospelovRamani2026,BoseEtAl2026,DiMauroShaikh2026}. These results motivate treating the $350\keV$ point below as a reference scale for the ultraviolet cancellation rather than as an established full-density thermal-Higgsino signal.

The recoil also admits alternatives that do not use the Higgsino weak current. Examples include exothermic scattering, elastic spin-dependent dark matter, inelastic electroweak multiplets, dark-photon excitation, pseudoscalar exchange, fermionic absorption and magnetic inelastic composites~\cite{BaerBarger2026,ElahiSchwaller2026,SmirnovGriffithBeacom2026,Yamashita2026,Unwin2026,LouLu2026,AsadiEtAl2026}. Our recent study of the LZ-favored elastic isovector response shows that resolved xenon nuclear interference produces different target and annual-modulation scaling from dark-sector excitation~\cite{KhanNuclear2026}. The recoil energy alone therefore does not determine the microscopic origin.

We work with the supersymmetric $331$ model obtained from the $SU(6)$ construction of Ref.~\cite{KhanLi2026}. The symmetry breaking proceeds as
\begin{equation}
SU(6)\longrightarrow SU(3)_C\times SU(3)_L\times U(1)_X
\longrightarrow G_{\rm SM}.
\label{eq:chain}
\end{equation}
The parent work specifies the $U(1)_X$ normalization, anomaly-free chiral matter, the two-pair electroweak Higgs sector, the $\mathbf6+\overline{\mathbf6}$ fields responsible for $331$ breaking, and complete unified messenger multiplets. The inverse-seesaw, unification and inflationary sectors of that model were studied in Ref.~\cite{KhanLi2026}; they are not repeated here. The present calculation uses only the field content, soft structure and symmetry breaking needed for the LZ-motivated electroweakino sector.

This differs from a direct GUT-scale bino--wino analysis. Above $v_{331}$ there is no elementary bino. Hypercharge is obtained only after $SU(3)_L\times U(1)_X$ breaking, so the cancellation has to be written in the physical $(X,L,C)$ gaugino basis and then matched through the broken neutral sector. Moreover, the relevant $SU(3)_L$ gaugino mass is comparable with the broken-vector scale, and the neutral fermions in the $331$-breaking and Higgs multiplets participate in the threshold. We therefore diagonalize the neutral gauge--Higgs sector of the renormalizable model rather than assigning an effective bino mass above $v_{331}$.
Nonuniversal gaugino masses have also been studied through nonsinglet $F$ terms in $SU(5)$, $SO(10)$ and $E_6$~\cite{EllisGaugino1985,Drees1985,MartinNonUniversal2009,ChakraborttyMohantyRao2014,MillerMorais2014}. Their low-energy ratios do not account for the additional neutral mixing at a $331$ threshold. We first derive the invariant relation in the $(X,L,C)$ basis and subsequently calculate the neutral eigenvalues. We then examine a symmetry-compatible higher-dimensional operator that splits the existing vectorlike messenger multiplets in the presence of a separate supersymmetric mass spurion. Its scalar-mass, gauge-running and gravitino consequences are evaluated. The additional spurion and the heavy-threshold spectrum are specified assumptions, so neither the small splitting nor unification is claimed as a consequence of the group embedding alone.

Section~\ref{sec:model} summarizes the part of the supersymmetric 331 model needed below. Section~\ref{sec:sumrule} derives the gaugino sum rule and compares it with other GUT boundaries. Section~\ref{sec:fullneutral} treats the neutral threshold and the sub-MeV target locus. Sections~\ref{sec:ordering}--\ref{sec:radiative} discuss state ordering, soft-sector feedback, gravitino ordering and radiative retuning. Section~\ref{sec:phenom} places the result in the recent LZ literature. The appendices give the neutral-matrix entries, the complementary neutral block, the adjoint trace pattern and messenger Dynkin indices. The numerical approximations are stated at their points of use.

\section{The supersymmetric 331 model from the supersymmetric \texorpdfstring{$SU(6)$}{SU(6)} GUT model}\label{sec:model}
\subsection{Embedding, chiral matter and symmetry breaking}
We use the supersymmetric 331 model from the supersymmetric $SU(6)$ GUT model of Ref.~\cite{KhanLi2026}. The diagonal generator associated with $U(1)_X$ is normalized in the fundamental representation as
\begin{align}
T_X&=\frac{1}{2\sqrt3}\,\mathrm{diag}(-1,-1,-1,+1,+1,+1),
&\mathrm{Tr}\,T_X^2&=\frac12.
\label{eq:TX}
\end{align}
Writing $q_6=1/(2\sqrt3)$,
\begin{equation}
\mathbf6=(\mathbf3,\mathbf1,-q_6)\oplus(\mathbf1,\mathbf3,+q_6),
\label{eq:6branch}
\end{equation}
and the two-index antisymmetric representation decomposes as
\begin{equation}
\mathbf{15}=(\overline{\mathbf3},\mathbf1,-2q_6)
\oplus(\mathbf1,\overline{\mathbf3},+2q_6)
\oplus(\mathbf3,\mathbf3,0).
\label{eq:15branch}
\end{equation}
The electric charge is
\begin{equation}
Q=T_{3L}+\frac{1}{\sqrt3}T_{8L}+\frac{2}{\sqrt3}X,
\label{eq:Q}
\end{equation}
which fixes the normalization of the hypercharge matching. Each family is assigned to $\mathbf{15}_i+\overline{\mathbf6}_i+\overline{\mathbf6}'_i$ together with a gauge singlet $N_{s i}$; the cubic $SU(6)$ anomaly cancels family by family~\cite{KhanLi2026,Deppisch2016,Le2020}. The vectorlike Higgs and messenger sectors do not introduce a net cubic anomaly.

The complete model employs $Z_4^R\times Z_2^\nu\times Z_3^G$ selection rules and contains the renormalizable terms~\cite{KhanLi2026}
\begin{align}
W={}&\frac14y^u_{ij}15_i15_jH_{15}
+\sqrt2y^d_{ij}15_i\overline6_j\overline H_6\nonumber\\
&+\frac14y^\nu_{ij}\overline6'_i\overline6'_jH_{15}
+Y_{ij}\overline6'_i\Phi N_{s j}\nonumber\\
&+\kappa S\left(\Phi\bar\Phi-\frac{v_0^2}{2}\right)
+W_H+W_G+W_{\rm GM}.
\label{eq:Wfull}
\end{align}
The first line generates the quark/lepton and inverse-seesaw structures studied in Ref.~\cite{KhanLi2026}; those flavor observables are not refitted here. The terms relevant to the present electroweakino problem are
\begin{align}
W_H={}&\mu_{15}\bar H_{15}H_{15}+\mu_6\bar H_6H_6\nonumber\\
&+\lambda_H H_{15}\bar H_6\bar\Phi
+\bar\lambda_H\bar H_{15}H_6\Phi,
\label{eq:WH}\\
W_{\rm GM}={}&\sum_{A=1}^{2}\lambda_A X\Psi_A\bar\Psi_A,
\label{eq:WGM}\\[-2pt]
\langle X\rangle={}&M_{\rm GM}+\theta^2F_X.
\end{align}
The complete $\mathbf6+\overline{\mathbf6}$ messenger pairs preserve the relative one-loop crossing of the three $331$ gauge couplings, while shifting the common unified coupling~\cite{KhanLi2026}. In the reference construction $M_{\rm GM}=3\times10^{15}\GeV$ and $\Lambda_X\equiv F_X/M_{\rm GM}=3\times10^5\GeV$.

The $Z_4^R$ assignments of the parent model leave a $Z_2$ subgroup unbroken by the $331$ vacuum. The Standard Model fermions are even under this residual parity, whereas gauginos, Higgsinos and scalar superpartners are odd. The lightest neutral superpartner is therefore stable within the effective theory provided the additional supersymmetry-breaking sector respects the same residual symmetry. This condition is used below when discussing the gravitino ordering.

The adjoint vacuum
\begin{equation}
\langle A\rangle=\frac{v_A}{2\sqrt3}\,
\mathrm{diag}(-1,-1,-1,+1,+1,+1)
\label{eq:Avec}
\end{equation}
breaks $SU(6)$ to $G_{331}$, while the lower $D$-flat direction is
\begin{equation}
\langle\Phi^0\rangle=\langle\bar\Phi^0\rangle=\frac{\vthree}{\sqrt2},
\qquad \vthree=30\TeV.
\label{eq:v331}
\end{equation}
The $331$ scale and the heavy threshold structure are inherited from the supersymmetric $SU(6)$ GUT model rather than introduced as an independent low-energy completion~\cite{KhanLi2026}.

\subsection{Two-pair electroweak Higgs sector}
After $331$ breaking the two electroweak doublet pairs have
\begin{equation}
M_D^{(0)}=
\begin{pmatrix}
\mu_{15}&\lambda_H\vthree/\sqrt2\\
\bar\lambda_H\vthree/\sqrt2&\mu_6
\end{pmatrix}.
\label{eq:MD0}
\end{equation}
A light supersymmetric doublet pair requires $\det M_D^{(0)}=0$. With
\begin{equation}
(\mu_{15},\mu_6,\lambda_H)=(7,6,0.2)\TeV,
\end{equation}
we obtain $\bar\lambda_H=0.4667$ and a heavy singular value $M_{D_H}=14.18\TeV$, in agreement with the reference construction~\cite{KhanLi2026}. Denoting the normalized left and right null vectors by $u$ and $d$, the low-energy supersymmetric Higgsino mass is introduced only along the light direction,
\begin{equation}
M_D=M_D^{(0)}+\mu\,u d^T,
\qquad |\mu|=1.091\TeV.
\label{eq:MDmu}
\end{equation}
This avoids identifying the heavy doublet singular value with the MSSM-like $\mu$ parameter.

\subsection{Gauge and soft matching at \texorpdfstring{$v_{331}$}{v331}}
With $g_1=\sqrt{5/3}\,g_Y$, the gauge couplings obey~\cite{KhanLi2026,Le2020}
\begin{equation}
\alpha_1^{-1}=\frac15\left(4\alpha_X^{-1}+\alpha_L^{-1}\right),
\qquad \alpha_2=\alpha_L,
\qquad \alpha_3=\alpha_C.
\label{eq:gmatch}
\end{equation}
For a supersymmetric interval it is convenient to define
\begin{equation}
I_a\equiv\frac{M_a}{g_a^2},
\label{eq:I}
\end{equation}
which is one-loop invariant in the absence of thresholds~\cite{MartinVaughn1994}. In the soft-threshold limit,
\begin{equation}
\whI=\frac45I_X+\frac15I_L,
\qquad I_2=I_L,
\qquad I_3=I_C.
\label{eq:Imatch}
\end{equation}
Here $\whI$ is the MSSM-equivalent hypercharge combination; it is not a propagating bino invariant above $\vthree$.

The universal gauge-mediated benchmark of Ref.~\cite{KhanLi2026} gives a gauge-only $331$-breaking scalar mass $m_\Phi(\vthree)\simeq7.76\TeV$ and low-scale gaugino masses approximately $(0.873,1.70,4.93)\TeV$. Our independent equation-based implementation gives $m_\Phi(\vthree)=7.60\TeV$ when the same gauge-only approximation is used. This agreement is used as a numerical cross-check. The nonuniversal source required below is then added and the scalar RGE is reintegrated; the $7.6$--$7.8\TeV$ value is not retained as a fixed input.

\section{Gaugino cancellation and comparison with unified boundaries}\label{sec:sumrule}
For heavy electroweak gauginos, integrating out the bino and wino generates the leading neutral-Higgsino Majorana operator. At matching scale,
\begin{equation}
\dmzero^{(5)}\simeq m_Z^2\left(\frac{s_W^2}{M_1}+\frac{c_W^2}{M_2}\right),
\label{eq:leadinggap}
\end{equation}
up to an overall sign convention~\cite{NagataShirai2015,ChunJungPark2017,DuWang2026}. The dimension-five term cancels for
\begin{equation}
\frac{M_1}{M_2}=-\tan^2\theta_W
=-\frac35\frac{g_1^2}{g_2^2}.
\label{eq:mssmcancel}
\end{equation}
Combining Eqs.~\eqref{eq:Imatch} and \eqref{eq:mssmcancel} gives
\begin{equation}
\boxed{I_X=-I_L.}
\label{eq:IXIL}
\end{equation}
The cancellation is therefore a relation between two physical $331$ gaugino invariants; it is not imposed on an elementary bino above $v_{331}$.

A singlet plus adjoint contribution to the $SU(6)$ gauge kinetic function gives a minimal realization. Writing
\begin{equation}
(I_X,I_L,I_C)=I_S(1,1,1)+I_A(0,1,-1),
\label{eq:spurionpattern}
\end{equation}
and $\eta=I_A/I_S$, one obtains
\begin{align}
\frac{I_X}{I_L}&=\frac{1}{1+\eta},&
\frac{I_C}{I_L}&=\frac{1-\eta}{1+\eta},\\
\frac{\whI}{I_L}&=\frac15\left(1+\frac{4}{1+\eta}\right).
\label{eq:etaratio}
\end{align}
The leading cancellation occurs at $\eta=-2$, which also predicts $I_C/I_L=-3$. This colored correlation is absent when the low-energy bino and wino are treated as independent parameters. The singlet--adjoint ratio is a parameterization of a soft boundary, not by itself a derived $F$-term vacuum.

\begin{figure*}[!t]
\centering
\includegraphics[width=\textwidth]{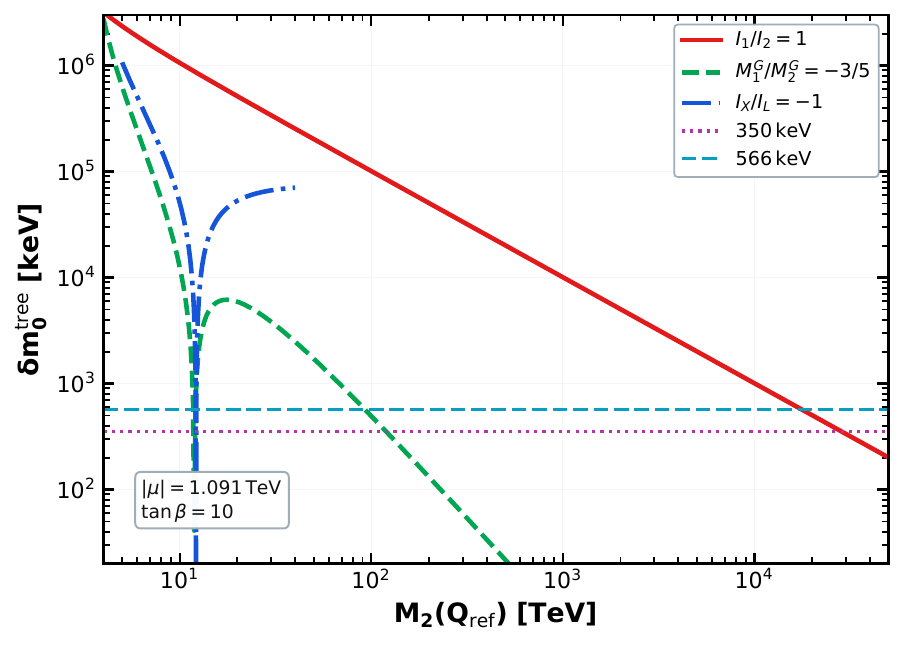}
\caption{Tree-level Higgsino splitting for three ultraviolet boundary structures. The curves correspond to same-sign universal MSSM invariants, the $SU(5)$ cancellation of Ref.~\cite{DuWang2026}, and the leading $I_X/I_L=-1$ relation of the supersymmetric 331 model before finite-threshold retuning. The horizontal lines mark the LZ and solar reference splittings.}
\label{fig:gut}
\end{figure*}

Figure~\ref{fig:gut} separates the effect of large gaugino masses from that of a cancellation. For equal-sign MSSM invariants the two terms in Eq.~\eqref{eq:leadinggap} add, and the tree-level splitting decreases approximately inversely with the gaugino scale. A $350\keV$ gap then requires $M_2\simeq2.9\times10^4\TeV$ in the displayed common-scale approximation. With opposite signs, the cancellation reduces the required scale: the $SU(5)$ boundary of Ref.~\cite{DuWang2026} has the tree-level roots near $11.82$, $12.01$ and $120.04\TeV$ shown in the figure. For the present $331$ embedding, $I_X/I_L=-1$ produces a comparably small gap at multi-TeV gaugino masses before finite-threshold retuning. Unlike an MSSM bino boundary, however, this line includes mixing with the broken $SU(3)_L\times U(1)_X$ sector. It is a comparison at a specified $\kappa$ and not an independent prediction of the $350\keV$ target. The simultaneous colored relation $I_C/I_L=-3$ arises only for the selected singlet--adjoint pattern~\cite{MartinNonUniversal2009,ChakraborttyMohantyRao2014,MillerMorais2014}.

\section{Neutral 331 threshold and state accounting}\label{sec:fullneutral}
\subsection{Hypercharge-connected ten-state block}
The soft-threshold expansion is not parametrically accurate when $M_L$ is comparable with the broken-vector mass. We therefore diagonalize directly the tree-level neutral states that communicate with the MSSM-like Higgsino through the displayed vacuum and renormalizable interactions. In the basis
\begin{equation}
\Psi^T=(\lambda_X,\lambda_8,\lambda_3,
\widetilde\Phi^0,\widetilde{\bar\Phi}^0,\widetilde S,
\widetilde H^u_{15},\widetilde H^u_6,
\widetilde H^d_{\bar{15}},\widetilde H^d_{\bar6}),
\label{eq:basis}
\end{equation}
the symmetric Majorana matrix is
\begin{equation}
{\cal M}_N=
\begin{pmatrix}
M_g&X_{g\Phi}&X_{gH}\\
X_{g\Phi}^T&M_\Phi&X_{\Phi H}\\
X_{gH}^T&X_{\Phi H}^T&M_H
\end{pmatrix},
\label{eq:MNblock}
\end{equation}
where $M_g=\mathrm{diag}(M_X,M_L,M_L)$. The $331$-breaking entries are
\begin{equation}
X_{g\Phi}=
\begin{pmatrix}
 g_Xq_6\vthree&-g_Xq_6\vthree&0\\
 -g_L\vthree/\sqrt3&+g_L\vthree/\sqrt3&0\\
0&0&0
\end{pmatrix},
\label{eq:XgPhi}
\end{equation}
and
\begin{equation}
M_\Phi=
\begin{pmatrix}
0&0&\kappa\vthree/\sqrt2\\
0&0&\kappa\vthree/\sqrt2\\
\kappa\vthree/\sqrt2&\kappa\vthree/\sqrt2&0
\end{pmatrix}.
\label{eq:MPhi}
\end{equation}
The electroweak gauge and superpotential mixing blocks are listed in Appendix~\ref{app:matrix}. No effective bino mass or effective bino--Higgsino coupling is inserted by hand.

Before electroweak breaking, the combinations $\widetilde\Phi_\pm=(\widetilde\Phi^0\pm\widetilde{\bar\Phi}^0)/\sqrt2$ separate the vector and singlet-like sectors. The $(\widetilde\Phi_+,\widetilde S)$ block is
\begin{equation}
{\cal M}_{+S}=\begin{pmatrix}0&\kappa\vthree\\kappa\vthree&0\end{pmatrix},
\label{eq:plusS}
\end{equation}
so the complete theory contains an additional pseudo-Dirac pair with mass $\kappa\vthree$. This state is absent in the MSSM neutralino matrix and gives the immediate ordering condition
\begin{equation}
\kappa>\kappa_{\rm LSP}\equiv\frac{|\mu|}{\vthree}=0.03637
\label{eq:kappalsp}
\end{equation}
if the Higgsino is to be the lightest neutral state at tree level.

\subsection{Complementary neutral vector-doublet block}
The ten-state system in Eq.~\eqref{eq:basis} is not the entire set of electrically neutral Weyl fermions carried by the displayed $SU(3)_L$ multiplets. The broken generators $T_6$ and $T_7$, the second components of $\Phi$ and $\bar\Phi$, and the neutral singlet components contained in $H_6$ and $\bar H_6$ form a second neutral block. The vacuum alignment $\langle\Phi\rangle=\langle\bar\Phi\rangle\propto(0,0,1)$ and the antisymmetric $SU(3)_L$ contractions make this block exactly decouple from Eq.~\eqref{eq:MNblock} at tree-level quadratic order in the displayed renormalizable sector. In the basis
\begin{equation}
\Psi_V^T=(\lambda_6,\lambda_7,\widetilde\Phi_2,
\widetilde{\bar\Phi}_2,\widetilde H_{6,3},
\widetilde{\bar H}_{6,3}),
\label{eq:Vbasis}
\end{equation}
its complex-symmetric matrix is given in Appendix~\ref{app:vectorblock}. At the lower $350\keV$ reference point its Takagi masses are
\begin{equation}
\{m_V\}/\TeV=\{6.000,6.000,8.834,8.834,20.961,20.961\}.
\label{eq:Vspectrum}
\end{equation}
Thus the neutral gauge--Higgs sector relevant to the electroweakino threshold contains sixteen states organized into the $10\oplus6$ blocks, and the complementary states remain above the Higgsino. The inverse-seesaw matter-neutrino system is a separate neutral block and does not mix with these states in the vacuum used here. Thus no additional neutral state in the displayed gauge--Higgs sector lies below the Higgsino at this point.

\subsection{Tree-level sub-MeV locus}
We adopt $Q_{\rm ref}=1\TeV$, $\tan\beta=10$, $|\mu|=1.091\TeV$ and use the gauge trajectory of the $SU(6)\to331$ construction. At $\vthree$, the equation-based evolution gives $g_X=0.4577$ and $g_L=0.6415$. For $M_2(Q_{\rm ref})=12\TeV$, the corresponding $SU(3)_L$ running mass is $M_L(\vthree)=12.13\TeV$, while the broken neutral-vector scale is $M_{V_0}=16.68\TeV$. Their proximity makes finite $331$ matching numerically relevant.

The neutral threshold depends on the $331$-breaking coupling $\kappa$ through the additional mass scale $\kappa v_{331}$. We therefore regard the target as a surface in $(M_2,\kappa,R_{XL})$ rather than as a single high-scale ratio. At the reference value $\kappa=0.07$, the hypercharge-connected block has a tree-level degeneracy at
\begin{equation}
\RXL\equiv\frac{I_X}{I_L}=-1.0026343.
\label{eq:Rc}
\end{equation}
The lower $350\keV$ branch is
\begin{align}
\RXL&=-1.0029981,\qquad \eta=-1.9970108,\nonumber\\
\frac{I_C}{I_L}&=-3.0059963,
\label{eq:R350}
\end{align}
with the opposite branch at $\RXL=-1.0022709$. The proximity to the leading $I_X=-I_L$ sum rule is therefore a property of the $331$ threshold, not a universal number. For example, at fixed $M_2(Q_{\rm ref})=12\TeV$ the lower $350\keV$ branch moves from $R_{XL}\simeq-0.873$ at $\kappa=0.05$ to $R_{XL}\simeq-1.070$ at $\kappa=0.12$. The target remains continuous throughout this range and is determined by the mass matrices rather than stored as a benchmark ratio.

At the lower solution the physical neutral spectrum is
\begin{align}
\{m_{\chi_i^0}\}/\TeV\simeq\{&1.090696,1.090696,2.10034,2.10042,\nonumber\\
&3.59738,12.1274,13.7240,14.1775,\nonumber\\
&14.1775,23.2546\}.
\label{eq:spectrum}
\end{align}
The two lightest states have a minimum Higgsino fraction $0.99957$ and a normalized off-diagonal $Z$ overlap $0.99957$. The inelastic weak current is therefore essentially the pure-Higgsino current. The singlet-like pair lies at $\simeq2.10\TeV$, above the dark-matter candidate.

\begin{figure*}[tbp]
\centering
\includegraphics[width=\textwidth]{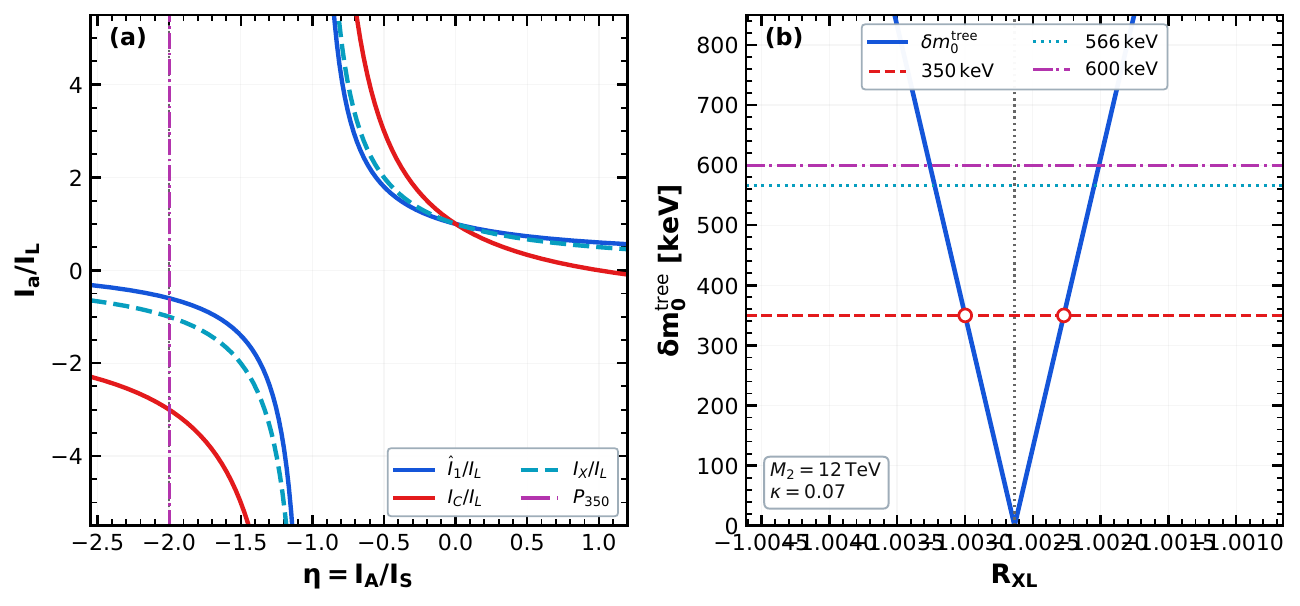}
\caption{Ultraviolet invariants and neutral-threshold matching. (a) Singlet--adjoint ratios from Eq.~\eqref{eq:etaratio}; $P_{350}$ denotes the lower $350\keV$ solution. (b) Tree-level Higgsino splitting from the hypercharge-connected block in Eq.~\eqref{eq:MNblock} at $M_2(Q_{\rm ref})=12\TeV$. The dotted vertical line gives the neutral degeneracy point.}
\label{fig:uvmatch}
\end{figure*}

\begin{figure*}[tbp]
\centering
\includegraphics[width=\textwidth]{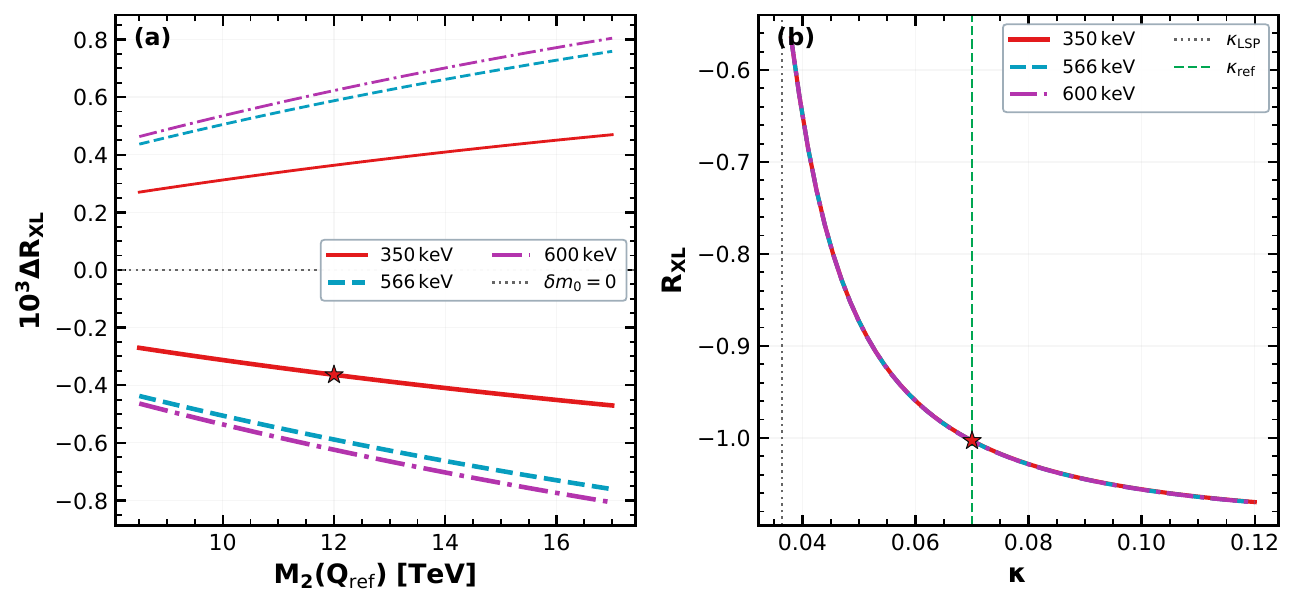}
\caption{Two projections of the tree-level target surface. (a) Displacement $\Delta R_{XL}$ from the degeneracy line as a function of $M_2(Q_{\rm ref})$ at $\kappa=0.07$. (b) Lower target branches as functions of $\kappa$ at $M_2(Q_{\rm ref})=12\TeV$. The star marks the reference $350\keV$ point.}
\label{fig:locus}
\end{figure*}

The two panels of Fig.~\ref{fig:uvmatch} illustrate distinct parts of the matching calculation. In panel (a), the Abelian and colored invariant ratios vary with the relative adjoint contribution $\eta$; the pole at $\eta=-1$ is caused by the normalization to $I_L$, not by a physical singularity in the gauge sector. The leading cancellation occurs at $\eta=-2$. Panel (b) shows the V-shaped tree-level splitting around the degeneracy at $R_{XL}=-1.0026343$. Since the gap is the difference of positive Takagi masses, each nonzero target has two nearby solutions, one on either side of the degeneracy. The $350\keV$ lower branch corresponds to $R_{XL}=-1.0029981$. Figure~\ref{fig:locus}(a) follows both branches with $M_2$, whereas panel (b) follows the lower roots as $\kappa$ changes. The variation in $R_{XL}$ is appreciable even when the required displacement from the local degeneracy line is small. This continuous target surface is a property of the calculated threshold, not a symmetry argument for its alignment.

\section{Neutral-state ordering and soft-sector backreaction}\label{sec:ordering}
\begin{figure*}[tbp]
\centering
\includegraphics[width=\textwidth]{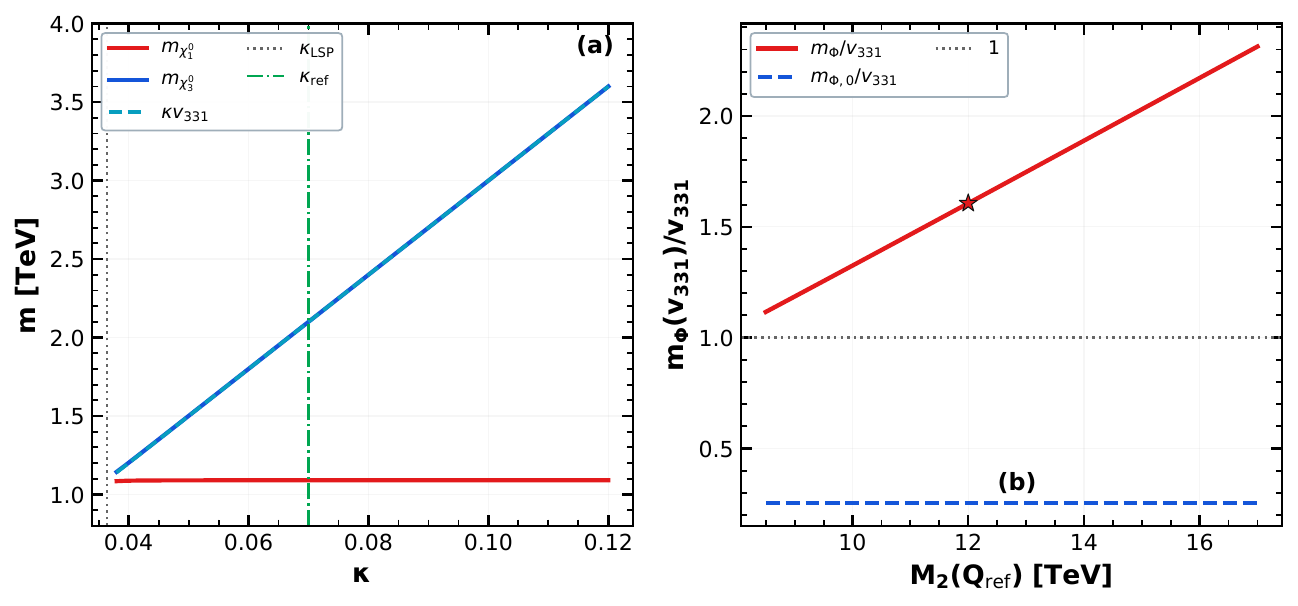}
\caption{Spectrum and soft-sector conditions along the LZ-motivated branch. (a) Light neutral masses on the lower $350\keV$ target as $\kappa$ varies. (b) Gauge/gaugino contribution to $m_\Phi(v_{331})/v_{331}$ along the same target as $M_2(Q_{\rm ref})$ varies. The dashed curves show the corresponding reference quantities defined in the text. Panel (b) uses the effective singlet--adjoint soft boundary; the distinct split-messenger scalar boundary is calculated in Sec.~\ref{sec:mediation}.}
\label{fig:consistency}
\end{figure*}

The $331$-breaking superpotential contains
\begin{equation}
W\supset \kappa S\left(\Phi\bar\Phi-\frac{v_0^2}{2}\right),
\end{equation}
so before electroweak breaking the $(\widetilde\Phi_+,\widetilde S)$ pair has mass $\kappa v_{331}$, Eq.~\eqref{eq:plusS}. A Higgsino dark-matter interpretation therefore requires
\begin{equation}
\kappa>\kappa_{\rm LSP}=\frac{|\mu|}{v_{331}}=0.03637.
\label{eq:kappalsp2}
\end{equation}
This condition concerns the spectrum rather than the cancellation itself. In Fig.~\ref{fig:consistency}(a), the nearly horizontal Higgsino mass is compared with the lightest additional neutral state along the $350\keV$ lower branch. As $\kappa$ approaches $|\mu|/v_{331}$, the singlet-like state approaches the Higgsino mass and can alter the identity of the lightest neutral particle. At the reference $\kappa=0.07$, it lies near $2.10\TeV$ and the lighter pseudo-Dirac pair remains dominantly Higgsino. The coupling also enters the inflationary sector of the parent construction~\cite{KhanLi2026}, but the inflationary constraints are not recalculated in this work.

The signed gaugino source also feeds back into the field that breaks the $331$ symmetry. Two ultraviolet implementations are distinguished below: the effective singlet--adjoint boundary of Eq.~\eqref{eq:spurionpattern} with the parent messenger scalar mass, and the explicitly split messenger model of Sec.~\ref{sec:mediation}. Their scalar boundaries are different. In the gauge/gaugino approximation used in the parent model,
\begin{align}
16\pi^2\frac{d m_\Phi^2}{d\ln Q}
&=-8\sum_a C_a^\Phi g_a^2|M_a|^2+\cdots,\nonumber\\
(C_X,C_L,C_C)&=\left(\frac1{12},\frac43,0\right).
\label{eq:mphiRGE}
\end{align}
We reconstruct the gauge-mediated boundary from the messenger equations and integrate Eq.~\eqref{eq:mphiRGE} rather than fixing the low-scale value. With the universal parent source, the code gives $m_\Phi(v_{331})=7.60\TeV$, close to the $7.76\TeV$ value reported in Ref.~\cite{KhanLi2026}. For the effective boundary in which the new singlet--adjoint source contributes to gaugino masses but not directly to the messenger scalar boundary, the lower $350\keV$ branch at $M_2(Q_{\rm ref})=12\TeV$ gives
\begin{equation}
m_\Phi(v_{331})=48.18\TeV.
\label{eq:mphiNew}
\end{equation}
The resulting soft contribution is large. It does not by itself remove the $331$ minimum because the stationarity condition can still be satisfied. In the minimal $D$-flat potential the stationarity equation is
\begin{equation}
v_0^2=v_{331}^2+\frac{2m_\Phi^2}{\kappa^2},
\qquad
\Delta_\Phi\equiv\frac{2m_\Phi^2}{\kappa^2v_{331}^2}.
\label{eq:DeltaPhi}
\end{equation}
For $\kappa=0.07$ this effective-source estimate has $\Delta_\Phi\simeq1.05\times10^3$. The split-messenger calculation of Sec.~\ref{sec:mediation} gives a larger gauge-only estimate, so the two soft boundaries are kept separate. Figure~\ref{fig:consistency}(b) shows the growth of the gauge/gaugino contribution to $m_\Phi$ with the wino reference mass while $R_{XL}$ is readjusted to remain on the $350\keV$ branch. The neutral cancellation thus does not suppress the scalar running, whose leading terms depend on $|M_a|^2$. Maintaining the chosen $331$ vacuum requires a compensating contribution in the scalar potential, and the quoted $\Delta_\Phi$ measures that adjustment within the minimal $D$-flat approximation.

\section{Nonuniversal soft source and gravitino ordering}\label{sec:mediation}
Nonuniversal gaugino masses can arise from nonsinglet $F$ terms in unified gauge kinetic functions~\cite{EllisGaugino1985,Drees1985,MartinNonUniversal2009}. The ratio in Eq.~\eqref{eq:etaratio}, however, does not determine an adjoint $F$ term or its alignment. There is also a discrete-symmetry restriction on the messenger construction. In the parent model the messenger pair has $Z_4^R$ charges $(0,0)$, the adjoint $A$ has charges $(R,q_G)=(0,1)$ under $Z_4^R\times Z_3^G$, and the superpotential carries $R=2$~\cite{KhanLi2026}. Consequently the seemingly simpler interaction $\bar\Psi A\Psi$ has $(R,q_G)=(0,1)$ and is forbidden. It cannot serve as the ultraviolet origin of the splitting.

A compatible effective operator can be written if a separate gauge-singlet mass spurion $Y$ is introduced with $(R,q_G)=(2,0)$. With $F_Y\simeq0$ and the adjoint vacuum of Eq.~\eqref{eq:Avec}, we take
\begin{equation}
W_{\rm mess}=\sum_{a=1}^{2}\bar\Psi_a\!\left[
X+\frac{\zeta}{M_*^3}YA^3\right]\!\Psi_a,
\label{eq:Wmesssplit}
\end{equation}
where $\langle X\rangle=M_{\rm GM}+\theta^2F_X$, $\langle Y\rangle=v_Y$ and the real coefficients refer to the representative CP-conserving benchmark. Both terms have $R=2$ and vanishing $Z_3^G$ charge: the cubic adjoint insertion is the first power allowed by the shaping symmetry. The additional singlet is not part of the field content specified in Ref.~\cite{KhanLi2026}. Its assumed scalar vacuum, negligible auxiliary expectation value and possible couplings to the rest of the theory are explicit ultraviolet conditions rather than predictions of the original model. In particular, a renormalizable $Y\bar\Psi\Psi$ term is also allowed by these charges. Its singlet contribution can be absorbed into the common messenger mass, but its coefficient is an additional ultraviolet input; the displayed benchmark sets it to zero. An illustrative supersymmetric singlet stabilization is discussed in Appendix~\ref{app:mess}.

Because the two $SU(3)$ blocks have opposite adjoint eigenvalues, $A^3$ acts as $\mp(q_6v_A)^3$ on their messenger components. The colored and left-handed signed holomorphic masses are therefore
\begin{equation}
{\cal M}_{C,L}=M_{\rm GM}(1\mp\xi),\qquad
\xi\equiv\zeta\frac{v_Y}{M_{\rm GM}}
\left(\frac{q_6v_A}{M_*}\right)^3,
\label{eq:messmasses}
\end{equation}
with physical threshold masses $|{\cal M}_{C,L}|$. Unlike a coupling of the form $Xf(A)\bar\Psi\Psi$, the independent $Y$-term changes the messenger supersymmetric masses without multiplying the $F_X$ insertion. Hence $F_C=F_L=F_X$ for $F_Y=F_A=0$, and the one-loop gaugino invariants can have different signs. For $|F_X|\ll|{\cal M}_{C,L}|^2$ the ordinary one-loop messenger graph gives
\begin{align}
I_C&=\frac{2F_X}{16\pi^2{\cal M}_C},&
I_L&=\frac{2F_X}{16\pi^2{\cal M}_L},\nonumber\\
I_X&=\frac12(I_C+I_L).
\label{eq:messinvariants}
\end{align}
The factor $1/2$ in the Abelian term follows from the normalized $U(1)_X$ index of each triplet, and it is derived in Appendix~\ref{app:mess}. Thus this pair of messenger representations enforces the one-loop relation $2I_X=I_C+I_L$ without requiring an independent adjoint $F$ term. For identical pairs,
\begin{align}
\frac{I_X}{I_L}&=\frac{1}{1-\xi},&
\frac{I_C}{I_L}&=\frac{1+\xi}{1-\xi}.
\label{eq:messratios}
\end{align}
The leading cancellation $I_X=-I_L$ is reached at $\xi=2$. It is still a selected messenger mass ratio rather than a consequence of the gauge group. In particular, the required sign reversal comes from the relative sign of the holomorphic messenger masses, not from a negative physical messenger mass.

At the lower $350\keV$ neutral-threshold branch of Eq.~\eqref{eq:R350}, the corresponding messenger input is $\xi=1.997011$. Fixing $I_L$ from the $12\TeV$ reference wino mass gives
\begin{align}
\frac{F_X}{M_{\rm GM}}&=6.974\times10^6\GeV,\nonumber\\
(|{\cal M}_C|,|{\cal M}_L|)&=(2.991,8.991)\times10^{15}\GeV.
\label{eq:messbenchmark}
\end{align}
The large hierarchy $|F_X|/|{\cal M}_{C,L}|^2\ll1$ validates the leading messenger expansion. The required $F_X/M_{\rm GM}$ is approximately $23$ times the universal gauge-mediated input of the parent study; the two sets of soft parameters cannot be interchanged. For illustration, taking $M_*=v_A$, $v_Y=10^{17}\GeV$ and $\lambda_X=1$ gives $\zeta\simeq2.49$ at the reference point. This realizes the signed masses at the level of the allowed effective operator, but does not derive the singlet vacuum, its auxiliary-field suppression or the chosen value of $\xi$.

\begin{figure*}[tbp]
\centering
\includegraphics[width=\textwidth]{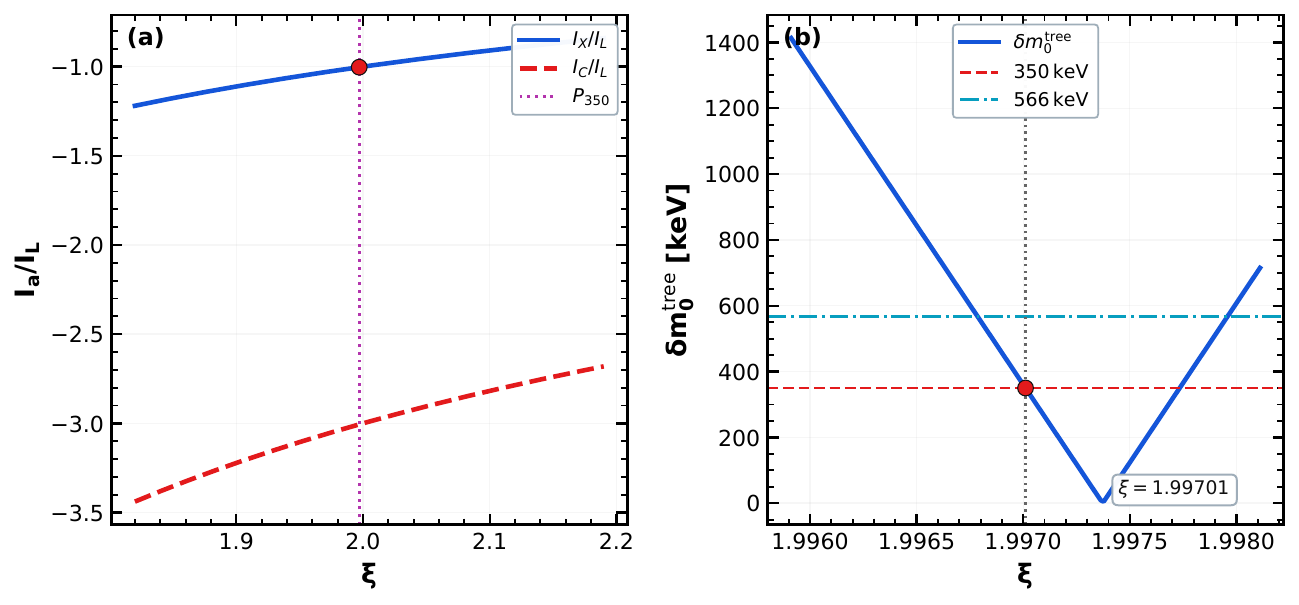}
\caption{Conditional symmetry-compatible effective messenger realization of the signed gaugino invariants. (a) One-loop ratios in Eq.~\eqref{eq:messratios}; the point marks the lower $350\keV$ reference solution. (b) Tree-level neutral splitting from the ten-state matrix with the messenger-induced ratio $R_{XL}=1/(1-\xi)$ and the original reference values of $M_2$ and $\kappa$. The narrow locus reflects the cancellation sensitivity.}
\label{fig:messuv}
\end{figure*}

A numerical derivative of the full ten-state neutral matrix yields
\begin{equation}
\Delta_R=\left|\frac{R_{XL}}{\delta m_0}
 \frac{\partial\delta m_0}{\partial R_{XL}}\right|\simeq2.76\times10^3,
\qquad
\Delta_\xi\simeq5.52\times10^3.
\label{eq:messsensitivity}
\end{equation}
These are local logarithmic sensitivities of a tree-level result, not statistical confidence intervals. Figure~\ref{fig:messuv}(a) shows how the two independent ratios vary as the adjoint interaction splits the messenger masses. In panel (b), inserting $R_{XL}=1/(1-\xi)$ in the full ten-state matrix produces a narrow V-shaped neutral gap. The $350\keV$ reference point lies close to the minimum, and a small departure in $\xi$ raises the gap into the solar-reference range. Writing a symmetry-compatible messenger operator makes the signed masses and colored correlation explicit, but it requires an additional singlet and does not determine $\xi$ dynamically; this residual tuning is expressed by $\Delta_\xi$.

\begin{figure*}[tbp]
\centering
\includegraphics[width=\textwidth]{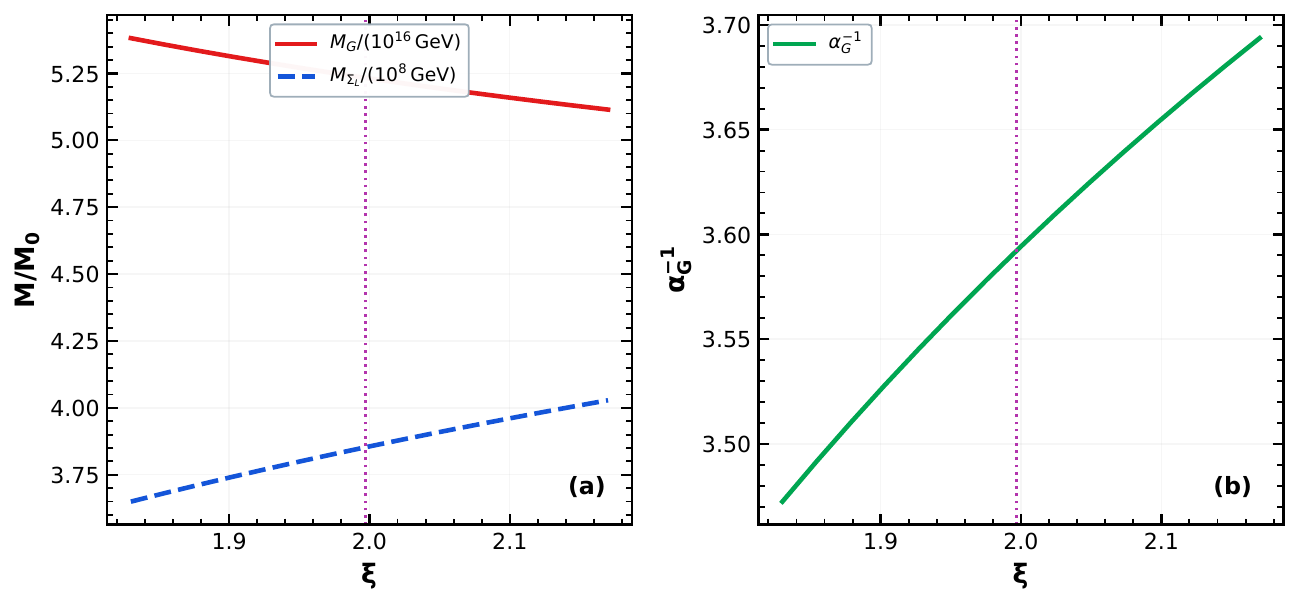}
\caption{Two-loop unification refit with adjoint-split messenger thresholds, holding the selected unified heavy-fragment ratios fixed. (a) Unified and left-adjoint threshold scales, with each curve divided by the scale shown in its legend. (b) Unified inverse coupling. The vertical line marks the neutral-splitting benchmark. The normalization $M_0$ in panel (a) is $10^{16}\,\mathrm{GeV}$ for $M_G$ and $10^8\,\mathrm{GeV}$ for $M_{\Sigma_L}$. The curves are conditional refits, not an uncertainty envelope from a computed heavy spectrum.}
\label{fig:messgut}
\end{figure*}

The split messenger masses also modify the relative gauge-coupling running. We repeat the two-loop $G_{331}$ evolution of Sec.~\ref{sec:model} with two colored messenger thresholds at $|{\cal M}_C|$ and two left-triplet thresholds at $|{\cal M}_L|$. With the same conditional unified heavy-fragment mass ratios $(0.70,1.35,0.85)$ as in the parent calculation and with $M_{\Sigma_L}$ refitted, the reference solution is
\begin{equation}
(M_G,M_{\Sigma_L},\alpha_G^{-1})=
(5.23\times10^{16},3.85\times10^8,3.59),
\label{eq:gutrefit}
\end{equation}
where masses are in GeV. The previous degenerate-messenger solution instead gave $M_G=4.49\times10^{16}\GeV$, $M_{\Sigma_L}=6.69\times10^8\GeV$ and $\alpha_G^{-1}=3.80$. Both calculations use the specified heavy-fragment thresholds as external inputs; neither determines the complete GUT Higgs spectrum or proton-decay widths. Figure~\ref{fig:messgut}(a) displays the compensating motion of $M_G$ and $M_{\Sigma_L}$ as $\xi$ is varied, with separate reference units defined in the legend. In panel (b) the unified inverse coupling changes on the same trajectory. The individual colored and left messenger contributions enter the running at $|{\cal M}_C|$ and $|{\cal M}_L|$, respectively, rather than at a common $M_{\rm GM}$. The retained crossing is obtained by refitting $M_{\Sigma_L}$ for the selected heavy-fragment spectrum. It establishes the sensitivity of the conditional solution, not threshold-independent unification. At the reference crossing $\alpha_G\simeq0.278$ and $g_G\simeq1.87$. The gauge-sector loop expansion remains below unity but is not highly suppressed; omitted Yukawa contributions, the incomplete heavy spectrum and higher orders limit the precision of this numerical crossing.

The gaugino source also generates a scalar soft boundary, unlike the additional gauge-kinetic $F$ term used for Fig.~\ref{fig:consistency}(b). At leading order in $F_X/{\cal M}_{C,L}^2$ the split messenger boundary is
\begin{align}
m_\Phi^2&=2\sum_{a=X,L}C_a^\Phi
 \left(\frac{g_a^2}{16\pi^2}\right)^2\Lambda_a^2,
\label{eq:splitsoft}\\
\Lambda_X^2&=\sum_{A=1}^2\frac12|F_X|^2
 \left(\frac1{|{\cal M}_C|^2}+\frac1{|{\cal M}_L|^2}\right),\nonumber\\
\Lambda_L^2&=\sum_{A=1}^2\frac{|F_X|^2}{|{\cal M}_L|^2},
\end{align}
with $C_X^\Phi=1/12$ and $C_L^\Phi=4/3$. Evaluating the couplings at $M_{\rm GM}$ gives a common-scale gauge-only boundary $m_\Phi\simeq56.5\TeV$; evolving only the gauge/gaugino contribution to $v_{331}$ gives an illustrative $m_\Phi\simeq62.6\TeV$. This estimate neglects the distinct messenger-scale matching in the scalar sector and the Yukawa and full soft-mass RGEs. In the minimal $D$-flat potential it corresponds to $\Delta_\Phi\simeq1.78\times10^3$. The adjoint-split completion thus increases, rather than eliminates, the vacuum sensitivity. A stable simultaneous solution of the complete scalar potential has not been demonstrated by this gauge-only computation.

The same messenger source changes the gravitino ordering. Taking $F_{\rm tot}\simeq F_X$ gives $m_{3/2}=F_X/(\sqrt3 M_{\rm Pl})\simeq4.96\TeV$; this is above the $1.091\TeV$ neutral Higgsino. The universal gauge-mediated point of Ref.~\cite{KhanLi2026}, with $F_X=9\times10^{20}\GeV^2$, instead has $m_{3/2}\simeq213\GeV$ and cannot yield a stable Higgsino LSP. The split source passes the simple ordering criterion
\begin{equation}
F_{\rm tot}>\sqrt3M_{\rm Pl}m_\chi\simeq4.60\times10^{21}\GeV^2.
\label{eq:Fmin}
\end{equation}
Nevertheless, Planck-suppressed soft terms, additional supersymmetry-breaking sectors, messenger stability and the separate $Y$-sector require a common ultraviolet construction before the full superpartner ordering can be considered established~\cite{GiudiceRattazzi1999,MeadeSeibergShih2009}.

\section{Radiative retuning}\label{sec:radiative}
The sub-MeV neutral gap is much smaller than ordinary electroweak radiative corrections. The exact solutions of Sec.~\ref{sec:fullneutral} are therefore matching-scale tree results, not pole-mass predictions. Below heavy scalar thresholds the Higgs--Higgsino--gaugino couplings run independently of the gauge couplings, and the bino-like and wino-like states are integrated out at different scales~\cite{GiudiceRomanino2004}. A complete pole calculation must combine sequential matching with neutralino and chargino self-energies in one renormalization scheme~\cite{Pierce1997,FritzscheHollik2002,Chatterjee2012,DuWang2026}.

The relevant question is whether finite corrections remove the solution or merely shift the high-scale invariant. Let
\begin{equation}
F(\RXL,p)=\dmzero(\RXL,p)-\dmzero^{\rm tar},
\end{equation}
where $p$ denotes running and threshold parameters. Near a simple root,
\begin{equation}
\delta\RXL=-\frac{\delta F}{\partial F/\partial\RXL}+{\cal O}(\delta F^2).
\label{eq:retune}
\end{equation}
Hence finite corrections continuously retune the invariant relation provided $\partial F/\partial\RXL\neq0$. The pointwise derivatives in Eq.~\eqref{eq:messsensitivity} quantify the price of this retuning. At the reference branch the numerical tree-level derivative is $|\partial\delta m_0/\partial\xi|\simeq0.967\,\mathrm{GeV}$. A hypothetical additive $0.1\MeV$ deformation would consequently displace $\xi$ by approximately $1.03\times10^{-4}$ in the local linear approximation. Larger deformations can cross the neutral-degeneracy cusp and need the full mass matrix. This is a sensitivity estimate, not a computed self-energy or a radiative correction. They do not establish a pole-mass prediction: a loop correction of ordinary electroweak size can exceed the target splitting, and the renormalized parameter must be adjusted in the same scheme as the self-energies.

\begin{figure*}[tbp]
\centering
\includegraphics[width=\textwidth]{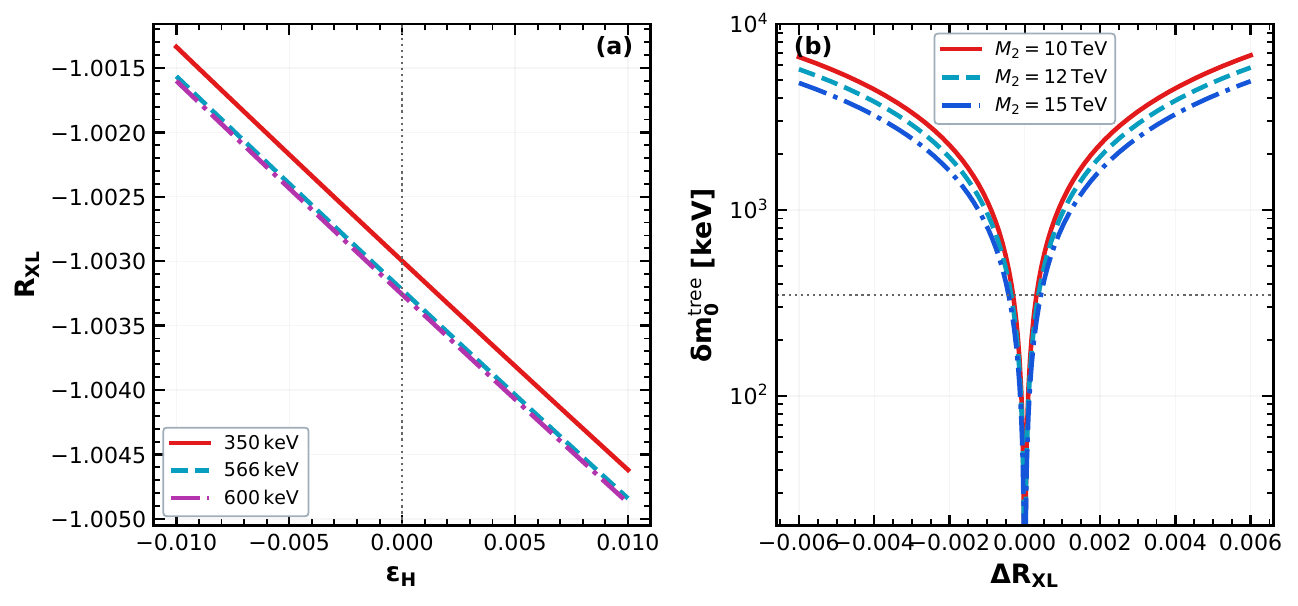}
\caption{Retuning diagnostics. (a) Lower target branch under the finite matching deformation of Eq.~\eqref{eq:epsH}. (b) Tree-level splitting as a function of displacement from the neutral degeneracy line for three values of $M_2(Q_{\rm ref})$.}
\label{fig:rad}
\end{figure*}

To display the local sensitivity without identifying it with a loop calculation, we deform the matched electroweak gauge mixing as
\begin{equation}
X_{gH}\rightarrow(1+\epsilon_H)X_{gH}
\label{eq:epsH}
\end{equation}
and rediagonalize the full matrix at each $\epsilon_H$. Figure~\ref{fig:rad}(a) shows that a percent-level change of the electroweak mixing entries displaces the required invariant ratio even though a sub-MeV tree-level target still has a solution. Panel (b) gives the local splitting as a function of displacement from the degeneracy line at three wino masses; the sharp minimum is the origin of the large sensitivity in Eq.~\eqref{eq:messsensitivity}. This exercise describes finite retuning of the tree-level mass matrix and is not a one-loop pole-mass calculation. In particular, the numerical root quoted at the reference scale does not have scheme-independent meaning at the keV level.

\section{LZ kinematics and phenomenological interpretation}\label{sec:phenom}
\begin{figure*}[tbp]
\centering
\includegraphics[width=\textwidth]{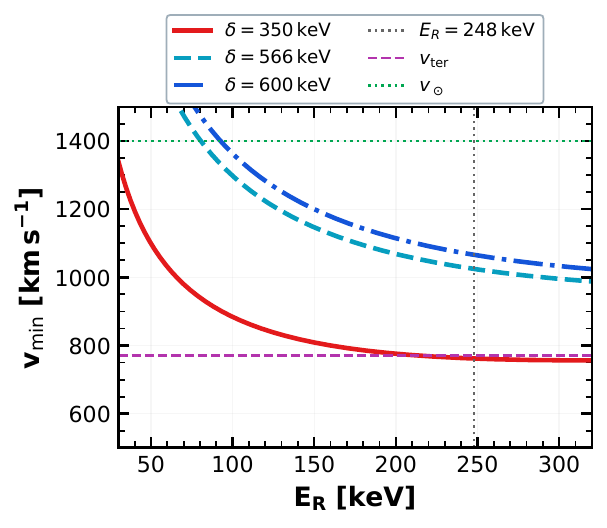}
\caption{Endothermic xenon kinematics. The vertical line marks the LZ recoil reference; $v_{\rm ter}$ and $v_\odot$ indicate representative terrestrial and solar velocity scales.}
\label{fig:vmin}
\end{figure*}

For endothermic scattering on a nucleus of mass $m_A$,
\begin{equation}
v_{\min}(E_R)=\frac{1}{\sqrt{2m_AE_R}}
\left(\frac{m_AE_R}{\mu_{\chi A}}+\dmzero\right),
\label{eq:vmin}
\end{equation}
where $\mu_{\chi A}$ is the reduced mass~\cite{TuckerSmithWeiner2001,Bramante2016}. Figure~\ref{fig:vmin} shows that the $350\keV$ curve reaches its smallest incident speed close to the high-energy xenon recoil region; at $E_R=248\keV$ the required speed is about $762\,\mathrm{km\,s^{-1}}$. The $566$ and $600\keV$ curves lie higher at the same recoil, beyond the illustrative terrestrial speed shown by the horizontal line. The separate $1400\,\mathrm{km\,s^{-1}}$ line represents a solar-interior reference speed, not a new terrestrial halo component~\cite{PospelovRamani2026}. These kinematic comparisons do not replace a detector likelihood or a solar-capture calculation.

\subsection{Leading-order Higgsino relic abundance and solar capture}\label{sec:relic}
\begin{figure*}[tbp]
\centering
\includegraphics[width=\textwidth]{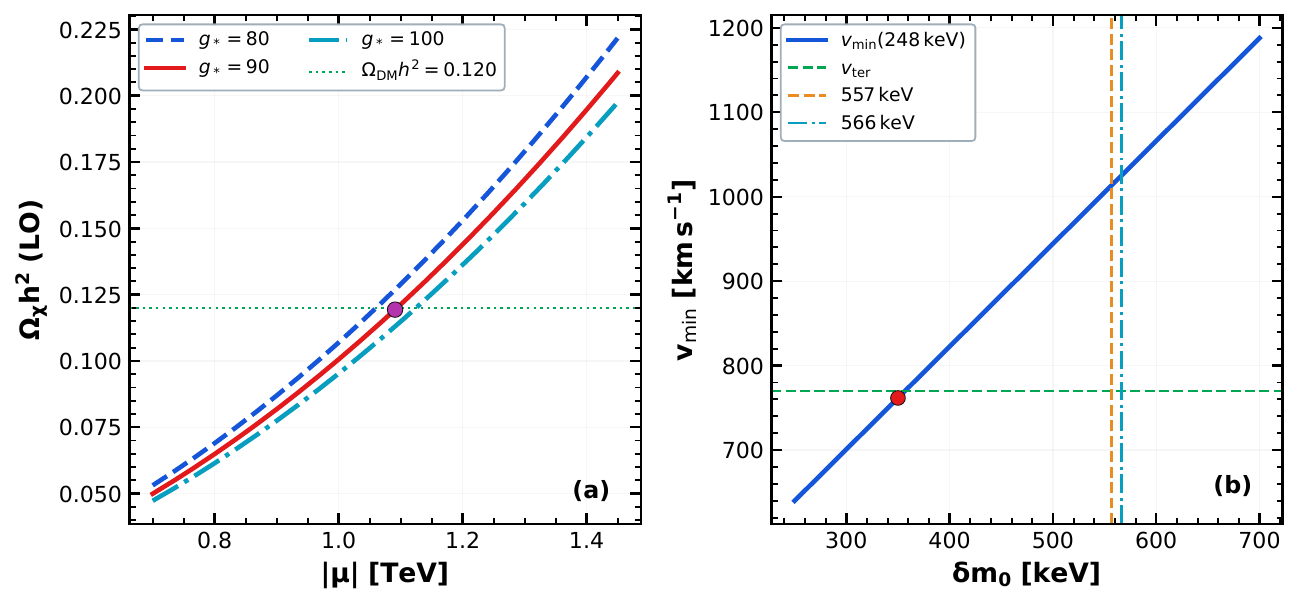}
\caption{Leading-order relic abundance and endothermic recoil kinematics. (a) Pure-Higgsino $s$-wave coannihilation freeze-out for the indicated constant $g_*$ values; the point is $|\mu|=1.091\TeV$. (b) Minimum incident speed for a $248\keV$ xenon recoil versus the neutral splitting. The vertical lines are published solar-capture reference bounds under their stated assumptions. The horizontal terrestrial speed is a reference used for orientation, not a detector or halo likelihood. The vertical solar markers are lower limits under the assumptions of the cited studies.}
\label{fig:therm}
\end{figure*}

We evaluate the leading pure-Higgsino coannihilation limit rather than identifying a mass close to the thermal value with an abundance calculation. For nearly degenerate two neutral Majorana fields and one charged Dirac fermion, the unbroken-electroweak $s$-wave rate is~\cite{SafdiTASI2023}
\begin{equation}
\langle\sigma_{\rm eff}v\rangle_{\rm LO}=
\frac{g_2^4}{512\pi m_\chi^2}
(21+3t_W^2+11t_W^4),\quad t_W=\tan\theta_W.
\label{eq:sigmarelic}
\end{equation}
With a radiation-dominated thermal history, $g_{\rm eff}=8$ and constant $g_*=90$, the sudden-freeze-out approximation gives
\begin{align}
x_f&=\ln\!\left[\frac{0.038\,g_{\rm eff}M_{\rm Pl}^{\rm(unred)}
 m_\chi\langle\sigma_{\rm eff}v\rangle_{\rm LO}}
 {\sqrt{g_*x_f}}\right],\nonumber\\
\Omega_\chi h^2&\simeq\frac{1.07\times10^9\GeV^{-1}\,x_f}
 {M_{\rm Pl}^{\rm(unred)}\sqrt{g_*}\langle\sigma_{\rm eff}v\rangle_{\rm LO}}.
\label{eq:omegarelic}
\end{align}
For $m_\chi=1.091\TeV$ the sudden-freeze-out calculation gives $x_f=25.85$, $T_f=42.2\GeV$ and $\Omega_\chi h^2\simeq0.119$. Figure~\ref{fig:therm}(a) compares the leading pure-Higgsino mass dependence at $g_*=80$, $90$ and $100$ with $\Omega_{\rm DM}h^2\simeq0.120$~\cite{Planck2018}. The mass dependence follows from the inverse-square annihilation rate, whereas the separation of the curves represents the fixed-$g_*$ approximation. Panel (b) holds the recoil energy at $248\keV$ and increases the excitation gap; the incident speed rises past the illustrative terrestrial reference before reaching the solar-capture splitting bounds. The abundance result is not a full $331$ freeze-out calculation: charged-state mixing and masses, subleading annihilation channels and Sommerfeld corrections are not included. The heavier neutral states are above $T_f$, but their effects on the charged spectrum have not been computed.

The normalized off-diagonal Higgsino $Z$ overlap is $0.99957$, so its weak coupling remains close to the pure-Higgsino value. Solar-capture studies excluding splittings below roughly $557$--$566\keV$ in their respective full-density assumptions consequently apply directly as reference constraints~\cite{BoseEtAl2026,PospelovRamani2026}. For the $248\keV$ xenon recoil the tree-level $350\keV$ point requires $v_{\min}\simeq762\,\mathrm{km\,s}^{-1}$, while $566\keV$ requires about $1025\,\mathrm{km\,s}^{-1}$. These numbers are kinematic, not a fit to the LZ likelihood or solar annihilation data. In particular, changing the messenger origin of the gaugino masses does not suppress the nearly unscreened weak current. Under standard full-density and halo assumptions the $350\keV$ benchmark does not furnish a consistent explanation of both the xenon event and the solar-neutrino limits.

The recent LZ literature distinguishes the required recoil kinematics from the origin of a small neutral mass gap. Endothermic recoil studies determine the preferred splitting and its sensitivity to the halo tail~\cite{SuYangYang2026,DiMauro2026,McCabe2026}. Higgsino analyses translate this scale into a fixed weak current and explore thermal, indirect-detection, high-scale-SUSY and collider consequences~\cite{FanReece2026,FreeseTheodosopoulos2026,WuZhangZhu2026,Yin2026,CheungKangKumar2026}. Ref.~\cite{DuWang2026} realizes the cancellation directly as an $SU(5)$ bino--wino boundary, whereas the GNMSSM uses Higgsino--singlino mixing and can retain multi-TeV gauginos~\cite{BisalCaoLi2026}. Minimal electroweak multiplets provide another inelastic realization~\cite{SmirnovGriffithBeacom2026}; pseudoscalar exchange, dark-photon excitation and fermionic absorption illustrate mechanisms whose recoil systematics do not rely on the Higgsino weak current~\cite{Unwin2026,Yamashita2026,LouLu2026}. A separate $SU(6)$-based GUT construction correlates the pseudo-Dirac splitting with a heavy $Z'$ and emphasizes future-collider tests~\cite{KotlarskiKowalskaSessolo2026}. In the supersymmetric 331 model from the supersymmetric $SU(6)$ GUT model, hypercharge appears only at the $331$ threshold. The ultraviolet relation is therefore written in the physical $(I_X,I_L,I_C)$ basis, while the supersymmetric $SU(6)$ GUT model fixes the breaking fields and messenger content used in the neutral threshold~\cite{KhanLi2026}.

For comparison, $SU(5)$, $SO(10)$ and $E_6$ nonsinglet $F$ terms can produce many representation-dependent MSSM gaugino ratios~\cite{MartinNonUniversal2009,ChakraborttyMohantyRao2014,MillerMorais2014}. Those constructions show that signed nonuniversal gaugino masses are compatible with supersymmetric unification. In the supersymmetric 331 model from the supersymmetric $SU(6)$ GUT model, the corresponding cancellation is $I_X\simeq-I_L$ above $v_{331}$, while the bino mass and bino--Higgsino interaction both arise after diagonalizing the broken $331$ sector. The minimal singlet--adjoint source simultaneously predicts $I_C/I_L\simeq-3$, so the colored boundary is correlated rather than independently adjustable.

The interpretation of the recoil remains separate from the ultraviolet calculation. Our recent study of the LZ-favored elastic isovector interaction finds a resolved xenon nuclear-interference feature and proposes target scaling and annual modulation as discriminants from dark-sector excitation~\cite{KhanNuclear2026}. Exothermic scattering, elastic spin-dependent dark matter, and magnetic inelastic composites provide further alternatives~\cite{BaerBarger2026,ElahiSchwaller2026,AsadiEtAl2026}. For the Higgsino considered here, the two light states retain an essentially unsuppressed off-diagonal $Z$ current, making solar capture a direct constraint. Pospelov and Ramani obtain a loop-informed reference near $566\keV$~\cite{PospelovRamani2026}; an independent analysis using Super-Kamiokande and IceCube data excludes splittings below roughly $557\keV$ for the stated thermal/nonthermal Higgsino assumptions~\cite{BoseEtAl2026}. Together with the LZ high-energy-sideband test~\cite{RoddSafdiSlatyerXu2026}, these results disfavor interpreting the $350\keV$ solution as an established full-density thermal-Higgsino signal. We use it as a reference point for the ultraviolet cancellation. The same neutral system admits larger reference splittings after per-mille shifts in $R_{XL}$, while elastic direct detection remains governed by the known loop-level Higgsino floor~\cite{Hisano2011,ChenHill2020,Martin2025}.

\section{Conclusions}\label{sec:conclusion}
We have studied pseudo-Dirac Higgsinos motivated by the LZ high-recoil event in the supersymmetric 331 model from the supersymmetric $SU(6)$ GUT model of Ref.~\cite{KhanLi2026}. The parent work already contains the unified matter, Higgs and messenger sectors; the present paper uses that structure only for the electroweakino problem. Since hypercharge appears after $331$ breaking, the leading MSSM bino--wino cancellation is replaced above $v_{331}$ by
$I_X=-I_L$. For the selected singlet--adjoint invariant pattern, the same condition gives $I_C/I_L=-3$. The conditional symmetry-compatible effective messenger operator gives this colored relation through the normalized triplet indices only when its mass ratio is chosen near the cancellation point; a new singlet spurion is required.

The finite threshold modifies this simple relation. The neutral gauge--Higgs fields of the renormalizable model separate into a hypercharge-connected ten-state block and a complementary six-state block. Direct diagonalization gives a continuous target surface in $(M_2,\kappa,R_{XL})$. For $M_2(Q_{\rm ref})=12\TeV$ and $\kappa=0.07$, the lower $350\keV$ branch occurs at $R_{XL}=-1.002998$; the two light states are more than $99.95\%$ Higgsino and retain the off-diagonal $Z$ current. The remaining neutral gauge--Higgs states lie above them. The decimal value is a reference matching result, not a universal GUT prediction.

The conditional messenger threshold produces $|{\cal M}_C|\simeq2.99\times10^{15}\GeV$, $|{\cal M}_L|\simeq8.99\times10^{15}\GeV$, and $F_X/M_{\rm GM}\simeq6.97\times10^6\GeV$ at the lower reference branch. Including the split messenger masses in the conditional two-loop gauge evolution changes the crossing to $M_G\simeq5.23\times10^{16}\GeV$, with the left-adjoint and unified heavy-fragment thresholds still treated as adjustable inputs. The local mass-splitting sensitivity is $\Delta_\xi\simeq5.52\times10^3$, which makes the residual alignment cost explicit.

The same parameter region imposes two further conditions. The $331$-breaking pair requires $\kappa v_{331}>|\mu|$ to keep the Higgsino as the lightest neutral state. In the messenger realization the gauge-induced scalar contribution reaches approximately $63\TeV$ in the stated leading estimate, increasing the adjustment needed to maintain the $331$ minimum. The enhanced spurion $F_X$ places the gravitino above the Higgsino in a simple supergravity estimate, unlike the original universal gauge-mediated point. The additional singlet vacuum, its auxiliary-field suppression, the full scalar vacuum and Planck-suppressed soft terms remain ultraviolet consistency conditions.

The sub-MeV values obtained here are tree-level matching quantities. Threshold corrections and neutralino--chargino self-energies shift the required high-scale alignment by amounts much larger than the target gap. The tree-level root persists under finite deformations, but a pole-level prediction requires sequential effective-field-theory matching in a specified renormalization scheme.

At leading order, standard Higgsino coannihilation gives $\Omega_\chi h^2\simeq0.119$ for the reference mass. This does not establish the model-specific abundance and does not resolve the solar-capture limits, because the relevant off-diagonal weak current is essentially unsuppressed.

The LZ event itself admits several interpretations. The $SU(5)$ signed-gaugino construction, Higgsino--singlino cancellation, electroweak multiplets and the separate $SU(6)$ collider-oriented model address the inelastic case in different ultraviolet settings~\cite{DuWang2026,BisalCaoLi2026,SmirnovGriffithBeacom2026,KotlarskiKowalskaSessolo2026}. Nuclear interference, exothermic scattering and other elastic or composite mechanisms have different target and halo dependences~\cite{KhanNuclear2026,BaerBarger2026,ElahiSchwaller2026,AsadiEtAl2026}. Solar-capture analyses disfavor the $350\keV$ full-density thermal-Higgsino interpretation under standard assumptions~\cite{PospelovRamani2026,BoseEtAl2026,DiMauroShaikh2026}. The present result is therefore a conditional statement about an explicitly parameterized messenger realization of a sub-MeV pseudo-Dirac Higgsino in the supersymmetric 331 model from the supersymmetric $SU(6)$ GUT model, with the experimental interpretation kept separate from the matching calculation.

\appendix
\section{Electroweak entries of the neutral matrix}\label{app:matrix}
The light null vectors of Eq.~\eqref{eq:MD0} satisfy
\begin{equation}
u^TM_D^{(0)}=0,
\qquad M_D^{(0)}d=0,
\qquad u^Tu=d^Td=1.
\end{equation}
For the reference parameters one may choose
\begin{align}
u&=(-0.81650,0.57735)^T,\nonumber\\
d&=(-0.51832,0.85519)^T.
\end{align}
up to independent overall signs. The electroweak vev vectors are $U_v=u\,v\sin\beta$ and $D_v=d\,v\cos\beta$.

The $X$, $T_8$, and $T_3$ charges of the two up-type neutral components are
\begin{align}
X_u&=(2q_6,q_6),\nonumber\\
T_{8u}&=\frac{1}{2\sqrt3}(-1,+1),\nonumber\\
T_{3u}&=(-1/2,-1/2).
\end{align}
The down-type conjugates carry the opposite $X$ and $T_8$ charges and $T_{3d}=(+1/2,+1/2)$. The gauge-mixing entries are generated component by component by
\begin{equation}
(X_{gH})_{ai}=g_aq_{ai}v_i.
\end{equation}
Electroweak symmetry breaking also differentiates Eq.~\eqref{eq:WH} with respect to $\Phi$ or $\bar\Phi$ and a doublet field, giving
\begin{align}
(X_{\Phi H})_{\bar\Phi,H_{15}^u}&=\lambda_H(D_v)_2/\sqrt2,\\
(X_{\Phi H})_{\bar\Phi,\bar H_6^d}&=\lambda_H(U_v)_1/\sqrt2,\\
(X_{\Phi H})_{\Phi,H_6^u}&=\bar\lambda_H(D_v)_1/\sqrt2,\\
(X_{\Phi H})_{\Phi,\bar H_{15}^d}&=\bar\lambda_H(U_v)_2/\sqrt2.
\end{align}
Together with Eqs.~\eqref{eq:XgPhi}, \eqref{eq:MPhi} and \eqref{eq:MDmu}, these expressions specify all entries of the hypercharge-connected block in Eq.~\eqref{eq:MNblock}.

\section{Complementary neutral vector-doublet block}\label{app:vectorblock}
Let $a=g_Lv_{331}/2$, $b_u=g_LU_6/2$, $b_d=g_LD_{\bar6}/2$, $c_u=\bar\lambda_H D_{\bar{15}}/\sqrt2$, and $c_d=\lambda_H U_{15}/\sqrt2$, where the electroweak vev components are those defined in Appendix~\ref{app:matrix}. In the basis of Eq.~\eqref{eq:Vbasis}, the remaining electrically neutral quadratic terms are
\begin{equation}
{\cal M}_V=
\begin{pmatrix}
M_L&0&-a&+a&-b_u&+b_d\\
0&M_L&-ia&-ia&+ib_u&+ib_d\\
-a&-ia&0&0&-c_u&0\\
+a&-ia&0&0&0&-c_d\\
-b_u&+ib_u&-c_u&0&0&\mu_6\\
+b_d&+ib_d&0&-c_d&\mu_6&0
\end{pmatrix}.
\label{eq:MVblock}
\end{equation}
The matrix is complex symmetric, so its physical Majorana masses are its Takagi singular values. The absence of entries linking Eq.~\eqref{eq:MVblock} to Eq.~\eqref{eq:MNblock} follows from the $(0,0,v_{331}/\sqrt2)$ vacuum alignment together with the antisymmetric $SU(3)_L$ contractions in $W_H$.

\section{Singlet--adjoint trace pattern}\label{app:trace}
Let the nonsinglet $F$ term be aligned with $T_X=q_6\,\mathrm{diag}(-\mathbf1_3,+\mathbf1_3)$. For an $SU(3)_C$ generator in the upper block,
\begin{equation}
\mathrm{Tr}(T_Xt_C^2)=-\frac{q_6}{2},
\end{equation}
whereas an $SU(3)_L$ generator in the lower block gives
\begin{equation}
\mathrm{Tr}(T_Xt_L^2)=+\frac{q_6}{2}.
\end{equation}
For the Abelian direction, $\mathrm{Tr}\,T_X^3=0$. Hence the adjoint contribution is proportional to $(0,+1,-1)$ in $(X,L,C)$, which yields Eq.~\eqref{eq:spurionpattern} after adding the singlet contribution.

\section{Dynkin indices and messenger matching}\label{app:mess}
For a $\mathbf6+\overline{\mathbf6}$ messenger pair, the upper colored triplet and lower left triplet have $U(1)_X$ charges $\mp q_6$, where $q_6=1/(2\sqrt3)$. Their contributions to the one-loop $331$ beta functions, in the order $(X,L,C)$, are
\begin{equation}
\Delta b_C=\left(\frac12,0,1\right),\qquad
\Delta b_L=\left(\frac12,1,0\right).
\label{eq:messindices}
\end{equation}
For example, two charged colored triplet multiplets give $2(3q_6^2)=1/2$ to the normalized Abelian index, while their combined $SU(3)_C$ Dynkin index is unity. In the small-$F_X$ limit, the holomorphic gauge-mediation formula is
\begin{equation}
\Delta I_a=\frac{F_X}{16\pi^2}\sum_r
 \frac{\Delta b_a^{(r)}}{{\cal M}_r}.
\label{eq:Ithreshold}
\end{equation}
Applying Eq.~\eqref{eq:Ithreshold} to two equal pairs gives Eq.~\eqref{eq:messinvariants}; the relation $2I_X=I_C+I_L$ holds before electroweak and $331$ threshold corrections.

For completeness, the operator in Eq.~\eqref{eq:Wmesssplit} is compatible with the displayed parent selection rules if $Y$ is assigned $(R,q_G,P_\nu)=(2,0,+)$. The messenger fields carry $R=0$, while $A$ has $(R,q_G)=(0,1)$; thus $X\bar\Psi\Psi$ and $YA^3\bar\Psi\Psi/M_*^3$ both have $(R,q_G)=(2,0)$. A lower-power insertion $YA^n$ with $n=1,2$ violates the shaping symmetry. One possible $F$-flat singlet module consists of $Y,\bar Y,T$, each with $R=2$ and $q_G=0$, together with
\begin{equation}
W_Y=\rho T(Y\bar Y-v_Y^2),
\end{equation}
whose global-supersymmetric branch has $\langle T\rangle=0$, $\langle Y\rangle\langle\bar Y\rangle=v_Y^2$, and $F_Y=F_{\bar Y}=F_T=0$. Their nonzero scalar expectations break $Z_4^R$ only to its $Z_2$ subgroup. This module supplies an illustrative supersymmetric spurion vacuum, not a complete common supergravity solution. The effect of soft terms, K\"ahler mixing and all other symmetry-allowed operators on $F_Y$ is not determined here. With $M_*=v_A$, $v_Y=10^{17}\GeV$ and $q_6=1/(2\sqrt3)$, Eq.~\eqref{eq:messmasses} gives $\zeta=2.4904$ for $\xi=1.9970108$. The small-expansion parameter for each adjoint eigenvalue is $q_6v_A/M_*=q_6$. The signed masses in Eq.~\eqref{eq:messmasses} retain the phases of the messenger determinant relevant for gaugino masses, while their absolute values enter gauge and scalar threshold functions. Below each messenger threshold, the invariants are evolved with the corresponding effective beta functions; the numerical code uses one-loop $M_a/g_a^2$ invariance in the gaugino sector and two-loop running for the gauge couplings.

To reproduce the conditional gauge-coupling comparison, the messenger increments of Eq.~\eqref{eq:messindices} are inserted at $|{\cal M}_C|$ and $|{\cal M}_L|$ rather than simultaneously at $M_{\rm GM}$. The other threshold assumptions and two-loop gauge-coefficient algorithm are left unchanged. The complete scalar soft evolution requires a separate threshold treatment of the split messenger multiplets, which is not supplied by Eq.~\eqref{eq:splitsoft} alone.

\section*{Data availability}
Available upon request to the corresponding author.

\section*{Declaration of competing interest}
The authors declare no competing financial or non-financial interests.
\begin{acknowledgments}
 The authors acknowledge that ChatGPT was used solely for language editing during the preparation of this manuscript, while the authors take full responsibility for the scientific content and final manuscript.

\end{acknowledgments}

\bibliography{references}
\end{document}